\documentclass[aps,pre,twocolumn,showpacs,superscriptaddress]{revtex4}

\usepackage[latin1]{inputenc}
\usepackage[english]{babel}
\usepackage{bm,graphicx,graphics,amsmath,amssymb,epsfig,color}

\usepackage[colorlinks,citecolor=blue,linkcolor=cyan]{hyperref}
\usepackage[usenames,dvipsnames,svgnames,table]{xcolor}

\def \be {\begin{equation}}
\def \ee {\end{equation}}
\def \beA {\begin{eqnarray}}
\def \eeA {\end{eqnarray}}

\def \der {\partial}

\def \Re  {\rm{Re}}
\def \Im  {\rm{Im}}
\def \average#1{\left\langle #1 \right\rangle}
\def \roundb#1{\left( #1 \right)}

\begin{document}
\title{Entropy production for complex Langevin equations}

\author{Simone Borlenghi} 
\affiliation{Department of Physics and Astronomy, Uppsala University, Box 516, SE-75120 Uppsala, Sweden.}
\author{Stefano Iubini}
\affiliation{ Dipartimento di Fisica e Astronomia, Universit\`a di Firenze, via G. Sansone 1 I-50019, Sesto Fiorentino, Italy}
\affiliation{Istituto Nazionale di Fisica Nucleare, Sezione di Firenze, via G. Sansone 1 I-50019, Sesto Fiorentino, Italy}
\author{Stefano Lepri}
\affiliation{Istituto dei Sistemi Complessi, Consiglio Nazionale delle Ricerche, 
Via Madonna del Piano 10 I-50019 Sesto Fiorentino, Italy.} 
\affiliation{Istituto Nazionale di Fisica Nucleare, Sezione di Firenze, via G. Sansone 1 I-50019, Sesto Fiorentino, Italy}
\author{Jonas Fransson}
\affiliation{Department of Physics and Astronomy, Uppsala University, Box 516, SE-75120 Uppsala, Sweden.}

\begin{abstract}
We study irreversible processes
for nonlinear oscillators networks described by complex-valued
Langevin equations that account for coupling to different thermo-chemical baths. 
Dissipation is introduced via non-Hermitian terms 
in the Hamiltonian of the model.
We apply the stochastic thermodynamics formalism to compute 
explicit expressions for the entropy production rates. We discuss 
in particular the non-equilibrium steady states of the 
network  characterised by a constant production rate of entropy and flows of 
energy and particle currents.
For two specific examples, a one-dimensional chain and a dimer, numerical calculations 
are presented. The role of asymmetric coupling among the oscillators on the 
entropy production is illustrated.
\end{abstract}

\pacs{05.60.-k, 05.70.Ln, 44.10.+}
\maketitle

\section{Introduction} %

Simple oscillator models allow one to tackle fundamental problems of non-equilibrium statistical mechanics \cite{BLRB00,lepri03,DHARREV,Basile08} and to 
study energy transport in systems that are ubiquitous in physics, chemistry, biology
and nanosciences \cite{balandin12,Lepri2016}. Examples include, but are not limited to, the dynamics of spin systems \cite{slavin09,borlenghi15b}, Bose-Einstein condensates, lasers, mechanical oscillators \cite{kevrekidis09} and photosynthetic reactions \cite{iubini15}.

A central issue is to identify the conditions under which a network of oscillators reaches thermal equilibrium, or is driven in a non-equilibrium steady state characterised by the propagation of coupled currents. One basic observable characterizing the state is the entropy production,
whose calculation in terms of the microscopic variables is the object of 
the present paper. In particular,  we address this issue  
using the language of Stochastic Thermodynamics (ST) \cite{crook99,searle02,seifert05,esposito06,deffner11,esposito12,seifert12}.
Within the ST framework, the out of equilibrium dynamics is described combining the Langevin and associated Fokker-Planck (FP) equations or a (quantum or classical)
master equation \cite{tome06,tome10,tome12,esposito10,esposito12,seifert12}. 
Those allow one to define the evolution of probability over the phase space and to derive consistent expressions for thermodynamic forces/flows and for entropy production
for states arbitrarily far from equilibrium.

Another issue that can be considered is the presence of asymmetric couplings in the system Hamiltonian.
Physical systems that can be described by asymmetrically coupled oscillators include magnetic materials with asymmetric exchange coupling
\cite{cheong07}, synthetic lattice gauge fields \cite{celi14}, transport in topological insulators \cite{rivas16} and parametrically driven oscillators \cite{salerno14}.
Here we discuss how, in a network of coupled oscillators, detailed balance can be broken either by the presence of thermal baths at different temperatures and chemical potential 
or by an \emph{anti-Hermitian coupling} among the oscillators.  
%
The use of anti-Hermitian Hamiltonians to describe 
phenomenologically irreversibility both in classical and quantum systems has been widely investigated \cite{dekker81,rajeev07,rotter09}.
Here we move a step forward by \emph{quantifying} irreversibility in those systems using the ST language. 

Although our formulation is completely general, we shall mostly refer to the dynamics of coupled nonlinear oscillators in the form of the discrete nonlinear Schr\"odinger equation (DNLS) \cite{Eilbeck1985,kevrekidis01,eilbeck03} 
whose off-equilibrium properties have received a certain attention recently \cite{iubini12,iubini13,borlenghi15b,Kulkarni2015,Mendl2015}. 
The spin-Josephson effect \cite{borlenghi15a}, the connection between gauge invariance and thermal transport \cite{borlenghi16a} and heat/spin rectification \cite{ren13,borlenghi14a,borlenghi14b} are a few of the effects within the DNLS field that
can be captured by the ST formalism.
One appealing feature of this class of models is the presence of two conserved quantities, namely energy and norm \cite{RASMUSSEN2000,iubini12,iubini13} that give rise to 
coupled transport effects between the associated currents \cite{iubini12}. This constitutes a further element of novelty that has not yet been considered in the existing
literature.

The remainder of the paper is organised as follows. In Sec. I we describe the dynamics of a network of complex Langevin equations, and we introduce the associated Fokker-Planck (FP) equation. 
In Sec. II we derive the entropy flow and entropy production for this system, and in Sec. III we identify the adiabatic and non-adiabatic components of entropy production. In Sec. IV we show the link between heat  and entropy flows and report simulations for 
the specific case of a DNLS chain with boundary thermostats. 
In Sec. V we discuss example of the dimer, the simplest realisation of the DNLS consisting of only two coupled oscillators. We present some numerical simulations that elucidate its 
off-equilibrium dynamics. Finally, in Sec. V we conclude the work and summarize the main results.

\section{Stochastic network model}%
\begin{figure}
\begin{center}
\includegraphics[width=6.0cm]{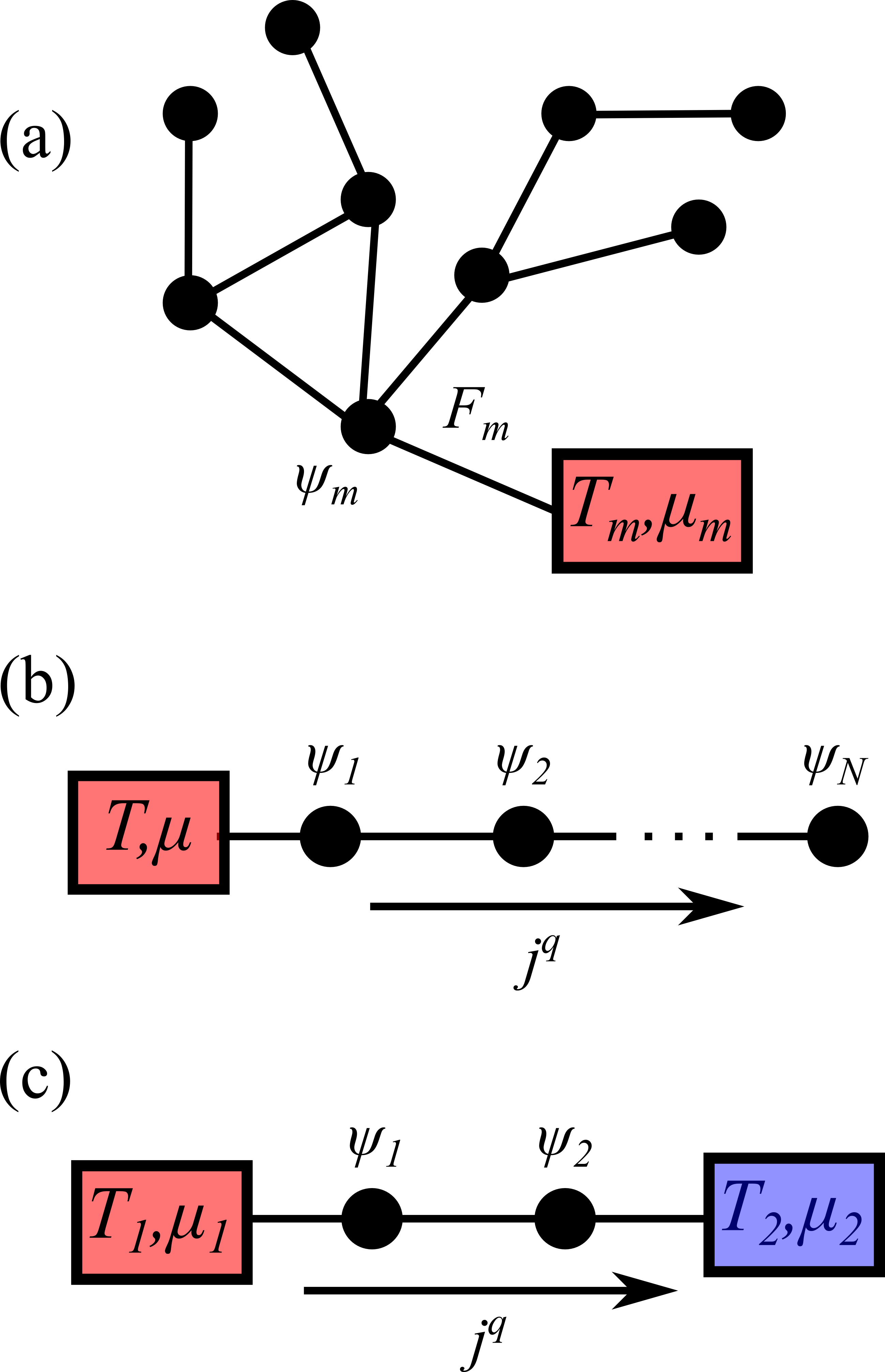}
\caption{(Color online) a) Network of nonlinear oscillators, where $\psi_m$ is the local oscillator amplitude and $F_m$ the force that specifies the geometry of the system.
Each site can be coupled to a thermal bath with temperature $T_m$ 
and chemical potential $\mu_m$. b) DNLS chain with the first site connected to a bath
with temperature $T$ and chemical potential $\mu$. c) Schr\"odinger dimer, consisting of two coupled oscillators connected to two thermal baths with different
temperature $T_m$ and chemical potential $\mu_m$, $m=1,2$.
}
\label{fig:figure0}
\end{center}
\end{figure}

Let us consider a network, where the dynamics of each of the $m=1,...,M$ nodes is described by the following Langevin equations (see the sketch in Fig.\ref{fig:figure0}(a))
\be\label{eq:langevin1}
\dot{\psi}_m=F_m+\xi_m\quad.
\ee
Here the dot indicates time derivative and $\psi_m=\sqrt{p_m(t)}e^{i\phi_m(t)}$ is a complex
oscillator amplitude. The force $F_m$ is an arbitrary function of the 
$\psi$s and their complex conjugate. 
We assume that both the coupling between the $\psi$s and the local forcing and damping are contained in the definition of $F$.
The white noises $\xi_m$, which model the stochastic baths, are complex Gaussian random processes with zero average and correlation  
\be\label{eq:bath}
\average{\xi_m(t)\xi^*_n(t^\prime)}=D_m\delta_{mn}\delta(t-t^\prime).
\ee

Here $D_m=\alpha_m T_m$ is the diffusion constant,  with $\alpha_m$ the damping rate and $T_m$ the temperature of bath $m$. 
Eq.(\ref{eq:langevin1}) is a general model that describes a multitude of systems encountered in physics, chemistry and biology.

Throughout the paper we adopt the following conventions: we set the Boltzmann constant $k_B$ equals to one. Vectors and matrices are written in
plain text, while their component are denoted by the $m$ and $n$ subscripts. 

We define the Wirtinger derivatives as 

\be\label{eq:wirt}
\der_m\equiv\frac{\der}{\der\psi_m}=\frac{1}{2}\roundb{\frac{\der}{\der x_m}-i\frac{\der}{\der y_m}},
\ee
with $\psi_m=x_m+iy_m$ and $\der_m^*=\frac{\der}{\der \psi_m^*}$ its complex conjugated. The variables $(\psi_m,i\psi_m^*)$ are canonically conjugate.
The total forces $F_m=F_m^I+F_m^R$ are the sum of dissipative (or irreversible, $I$) and conservative (or reversible, $R$)   components. 
Those are given by the derivatives $F_m^{I/R}=i\der_m^*\mathcal{H}^{I/R}$ of anti-Hermitian/Hermitian Hamiltonians $\mathcal{H}^{I/R}$, respectively.
The latter have opposite parity  under the time reversal transformation and the total Hamiltonian $\mathcal{H}$ is defined as $\mathcal{H}=\mathcal{H}^{I}+\mathcal{H}^{R}$.

The Fokker-Planck (FP) equation associated to Eq.(\ref{eq:langevin1}) reads \cite{haken69,bung10}
\be\label{eq:fp}
\dot{P}=\sum_m\left[ -\der_m(F_mP)-\der_m^*(F_m^*P)+2D_m\der_m\der_m^*P\right].
\ee
Eq.(\ref{eq:fp}) gives the evolution of the probability $P$ to find the system in the configuration $(\psi_1,...,\psi_M,\psi_1^*,...,\psi_M^*)$ at time $t$.
Following Refs.\cite{spinney12,landi13}, we define the irreversible and reversible probability currents 
\beA
\mathcal{J}^{I}_m &=&F^{I}_m P-D_m\der^*_m P,\label{eq:ji}\\
\mathcal{J}^{R}_m &=&F^{R}_m P\label{eq:jr},
\eeA
with  $\mathcal{J}_m=\mathcal{J}_m^I+\mathcal{J}_m^R$ and $\mathcal{J}_m^*$ its complex conjugated. In terms of those currents the FP equation Eq.(\ref{eq:fp}) assumes the form of a continuity equation:
\be\label{eq:cont}
\dot{P}=\sum_m \left(-\der_m\mathcal{J}_m-\der_m^*\mathcal{J}_m^*\right).
\ee
The steady state corresponds to $\dot{P}=0$, while thermal equilibrium corresponds to $\mathcal{J}_m=\mathcal{J}_m^*=0$.

The average of an arbitrary function $f$ of the observables is expressed by means of $P$ as $\average{f}=\int f P dx$, where $dx=(\frac{i}{2})^N\prod_{m=1}^N d\psi_m\wedge d\psi_m^*$ is the 
phase space volume element. Note that this average is equivalent to ensemble-average of Eq.(\ref{eq:langevin1}) over different realisations of the stochastic processes. 
As usual \cite{tome06,esposito12}, we consider the case where the probability currents and the thermodynamical forces
vanish at infinity, so that the cross terms in the integration by part can be discarded.

\section{Entropy flow and entropy production}%

The entropy flow $\Phi$ and entropy production $\Pi$ are obtained starting from the definition of phase space entropy 
\be\label{eq:entropy0}
S=-\average{\log P}\equiv -\int P \log P dx
\ee
and computing its time derivative by means of Eq.(\ref{eq:fp}): 
\be\label{eq:entropy1}
\dot{S} =\int\sum_m(\der_m\mathcal{J}_m+\der^*_m\mathcal{J}^*_m)\ln Pdx.
\ee 
Upon integrating by parts, using Eqs.(\ref{eq:ji}), (\ref{eq:jr}) and (\ref{eq:cont}) and assuming that the reversible forces have zero divergence \cite{tome06,tome10}, Eq.(\ref{eq:entropy1}) 
becomes 
\be\label{eq:entropy2}
\dot{S}=-2{\Re}\int\sum_m \mathcal{J}^I_m\frac{\partial_m P}{P}dx.
\ee
From Eqs. (\ref{eq:ji}) one has 
\be\label{eq:entropy3}
\frac{\der_mP}{P}=\der_m\ln P = \frac{1}{D_m}(F^{I*}_m-\mathcal{J}^{I*}_m/P)
\ee
and $\der_{m}^*\ln P$ its complex conjugate. Substituting this into Eq.(\ref{eq:entropy2}) gives
\be\label{eq:entropy4}
\dot{S}=-2{\Re}\int\sum_m \mathcal{J}^I_m\frac{F_m^I}{D_m}dx+\int\sum_m \frac{|\mathcal{J}_m^I|^2}{D_mP}dx.
\ee
The two terms in Eq.(\ref{eq:entropy4}) correspond respectively to minus the entropy flow $\Phi$ from the system to the environment 
and entropy production $\Pi$.
Note in particular that $\Phi$ has the usual form of products between probability fluxes $\mathcal{J}_m$ and thermodynamical forces $F_m^*/D_m$ and 
that $\Pi$ is positive-definite.
In non-equilibrium stationary states, one has $\dot{S}=-\Phi+\Pi=0$, so that $\Phi=\Pi$. These quantities are both zero only at thermal equilibrium.
Upon using Eq.(\ref{eq:ji}), integrating by parts and substituting the integrals over $P$ with ensemble average, the total entropy flow becomes

\be\label{eq:entropy5}
\Phi= \sum_m \Phi_m=\sum_m \left[2\frac{\average{|F^I_m|^2}}{D_m}+2{\Re}\average{\der_mF^I_m}\right]
\ee
where $\Phi_m$ is the entropy flow on site $m$.
Note that Eq.(\ref{eq:entropy5}) is the generalization of the expression given in Ref.\cite{tome06} to the case where forces are complex-valued. 

Before concluding the section, let us briefly discuss the more general case of non-stationary conditions.
To this aim, it is useful to separate the entropy production into adiabatic and non-adiabatic components, which correspond respectively to steady and non steady 
states [\onlinecite{esposito10}]. Upon indicating with superscript $s$ the steady state probability $P^s$ and fluxes $\mathcal{J}^s$, one writes the steady state FP equation as
\be\label{eq:fpstationary}
\dot{P}^s=\sum_m\left[-\der_m\mathcal{J}_m^s-\der_m^*\mathcal{J}_{m}^{*s}\right]\equiv 0.
\ee
By using Eqs.(\ref{eq:ji}) and (\ref{eq:jr}), it is convenient to define the following quantity
\be\label{eq:lambda1}
\Lambda_m\equiv\frac{\mathcal{J}_m}{P}-\frac{\mathcal{J}_m^{s}}{P^s} =-D_m\der^*_m\ln{\frac{P}{P^s}}.
\ee
Eq.(\ref{eq:lambda1}) defines the discrepancy between a stationary and non-stationary state.
By inserting $\Lambda_m$ into the definition of entropy production Eq.(\ref{eq:entropy2}) and integrating by parts, one can show that the latter splits
into the sum $\Pi=\Pi^a+\Pi^{na}$ of two parts which are respectively the adiabatic and non adiabatic components:
\beA
\Pi^{a} &=& 2\int\sum_m\frac{P}{D_m}\frac{|\mathcal{J}_m^s|^2}{P_s^2} dx=2\sum_m \average{\frac{|\mathcal{J}_m^s|^2}{D_mP_s^2}}\label{eq:pa}\\
\Pi^{na} &=& 2\int\sum_m\frac{P}{D_m}|\Lambda_m|^2dx =2\sum_m \average{\frac{|\Lambda_m|^2}{D_m}}\label{eq:pna}
\eeA
The adiabatic component corresponds to non-equilibrium steady state, obtained for example connecting the system to baths at different constant
temperature. On the other hand, the non-adiabatic component corresponds to non stationary states, obtained by applying a time dependent
driving to the system.

\section{Steady state heat flow}%

Let us return to the stationary case and consider the relation between the entropy flux $\Phi$ derived in the previous section and the heat flow.
For clarity, we specialize to the relevant case of DNLS oscillators in contact with boundary reservoirs \cite{iubini13}.
In particular, we consider the geometry sketched in Fig.\ref{fig:figure0}(b), where the first site of the chain is in contact with a reservoir 
on the left at temperature $T$ and chemical potential $\mu$ and with the rest of the chain on the right.
This setup is described by the following Hamiltonians~\cite{borlenghi15b}
\begin{eqnarray}
 \mathcal{H}^{R}&=&\sum_m h_m,\\
 \mathcal{H}^{I}&=&i\alpha(h_1-\mu p_1),
\end{eqnarray}
where $h_m$ is the local energy yielding the  conservative forces  $F_m^R\delta_{mj} = i \partial_m^* h_j$. Analogously,  the  irreversible forces are 
$F_m^I \delta_{m1} = i \partial^*_m \mathcal{H}^{I} $. 
Let us now evaluate the variation of the local internal energy $u_1 = \langle h_1 - \mu p_1\rangle $ on a stationary state.

\beA
\dot u_1 &=& \frac{d}{dt} \int dx P(x) (h_1-\mu p_1) = \int dx P(x) (\dot h_1 - \mu \dot p_1)\nonumber\\
             &=& \frac{1}{i\alpha} \int  dx P(x) \dot{\mathcal{H}}^{I},
\eeA
where $\alpha\neq 0$ is assumed. Upon substituting the dissipative forces and using the anti-hermitianity of $\mathcal{H}^{I}$, one has

\beA
\dot u_1 &=& \frac{1}{i\alpha}\left \langle\sum_m \left( \frac{\partial {\mathcal{H}}^{I}}{\partial \psi_m^*}\dot\psi_m^*
             +\frac{\partial {\mathcal{H}}^{I}}{\partial \psi_m}\dot\psi_m \right)\right\rangle\nonumber\\ 
             &=& \frac{1}{i\alpha}\left \langle \left( -iF_1^I\dot\psi_1^* -iF_1^{I*}\dot\psi_1\right)\right\rangle.
\eeA
By inserting the equations of motion, Eq.~(\ref{eq:langevin1}),  the above equation becomes
\beA
\dot u_1 &=& -\frac{1}{\alpha} \left[ 2 \langle |F_1^I|^2\rangle + 2{\Re} \langle F_1^I \xi_1(t)\rangle   \right]\nonumber\\
              &&-\frac{2}{\alpha} {\Re}\langle F_1^I F_1^{*R} \rangle 
\eeA
assuming that $\langle F_m^I \xi_j \rangle= \alpha T\langle \partial_m F_m^I\rangle $ as in Refs. \cite{tome06,tome10}, one gets

\be\label{eq:udot}
\dot u_1 = - \Phi_1T - \frac{2}{\alpha} {\Re}\langle F_1^I F_1^{*R} \rangle  = - \Phi_1 T - j^q_1,
\ee
where $j^q_1$ is the heat flux on site $1$. Indeed, for a lattice site $m$ in contact with the reservoir, we have 

\beA
\label{eq:heat_f}
j_m^q-j_{m-1}^q&=&\frac{2}{\alpha} {\Re}\langle F_m^I F_m^{*R} \rangle  =-2{\Re} \left\langle  \frac{\partial h_m}{\partial \psi_m^*} F_m^{*R}  \right\rangle\nonumber\\
&&+2\mu {\Re}\left\langle  \frac{\partial p_m}{\partial \psi_m^*} F_m^{*R}  \right\rangle.
\eeA 
The first term of the right-hand side  corresponds to the energy flow difference $j^h_m-j^h_{m-1}$ while the second term is the particle flow 
difference $j_m^p-j_{m-1}^p$~\cite{borlenghi15b}
multiplied by the chemical potential. Therefore, we consistently obtain the relation $j_m^q= j_m^h-\mu j_m^p$~\cite{Lepri2016}.
%
Finally, since $\dot u_1 =0$ on a stationary state and $j_0^q=0$, we recover the basic thermodynamical relation $j^q_1 =-\Phi_1 T$.
\begin{figure}
\begin{center}
\includegraphics[width=8.0cm]{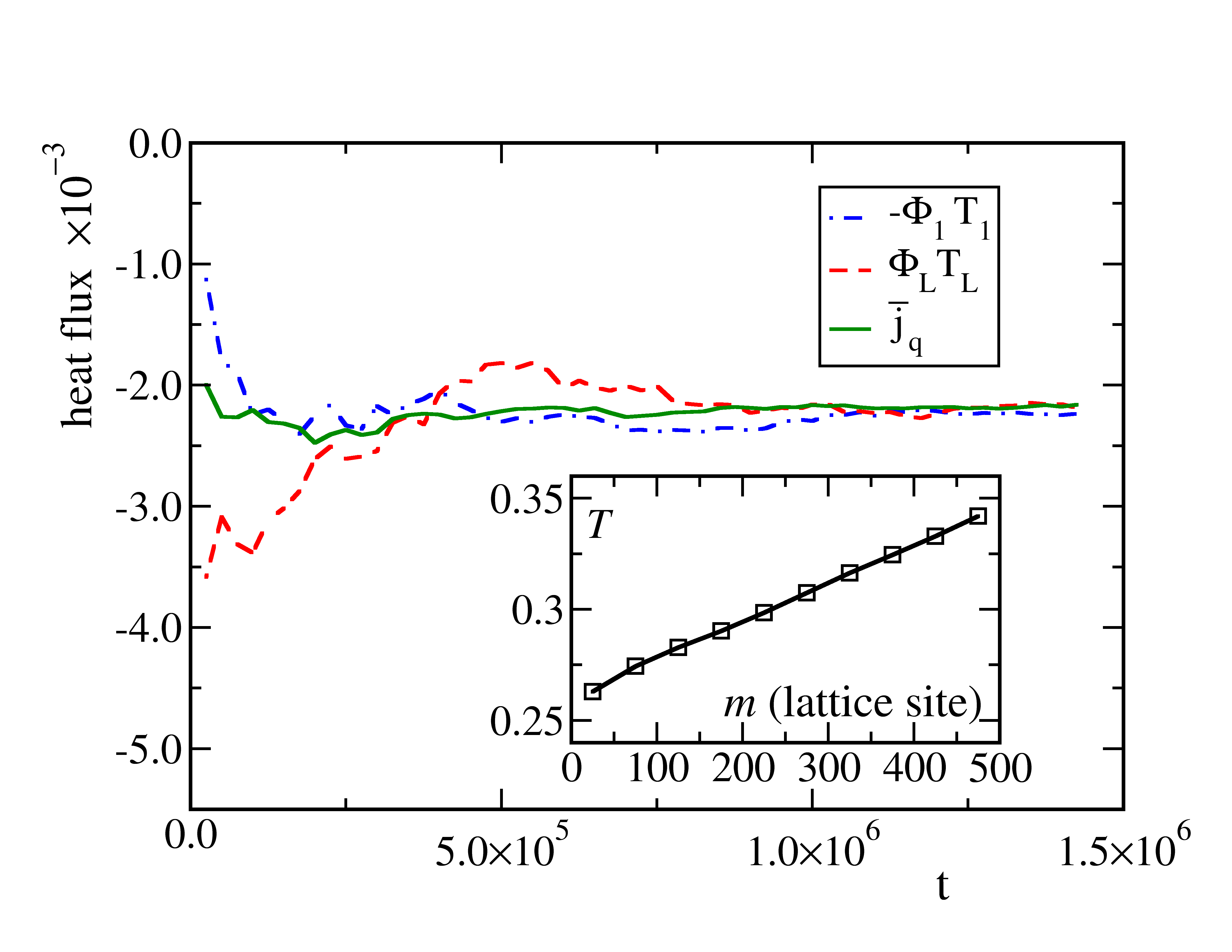}
\caption{(Color online) Heat-flux balance during the relaxation to a nonequilibrium stationary state of a DNLS chain with $L=500$ lattice sites.
The system is in contact with 
two boundary reservoirs at temperature $T_1=0.25$, 
$T_L=0.35$ and chemical potential $\mu_1=\mu_L=-1$, with couplings $\alpha_1=\alpha_L=0.05$. 
Blue (dot-dashed) and red (dashed) curves refer to the boundary heat flux computed from the entropy fluxes
$\Phi_{j}$, with $j=1,L$. The green (solid) line shows the behavior of the average heat flux $\overline{j_q}$ in the bulk, computed through Eq.~(\ref{eq:heat_f}).  
The inset shows the temperature profile measured in the stationary regime (see Ref.~\cite{iubini12} for computational details).
Simulations were performed with a 4-th order Runge-Kutta algorithm with a time step of $0.005$ model temporal units.
}
\label{fig:chain}
\end{center}
\end{figure}

The consistency of Eq.~(\ref{eq:udot}) has been tested numerically on a chain of $L$ DNLS oscillators in contact with two boundary
heat baths. The system Hamiltonian can be  explicitly written as 
\be
  \mathcal{H}^{R}=\sum_{m=1}^L \left( |\psi_m|^4 +\psi_m^*\psi_{m+1} + \psi_{m+1}^*\psi_m\right)
\ee
and  the heat baths are implemented as in Eq.~\ref{eq:langevin1} with $D_1=\alpha T_1$ and $D_2=\alpha T_2$. 
Assuming fixed boundary conditions $(\psi_0=\psi_{L+1}=0)$, the dissipative Hamiltonian reads~\cite{iubini13,borlenghi15b}
\beA
 \mathcal{H}^{I} &=& i\alpha(|\psi_1|^4+\psi_1^*\psi_2+\psi_2^*\psi_1-\mu_1 p_1)\nonumber\\
                 &&+i\alpha(|\psi_L|^4+\psi_L^*\psi_{L-1}+\psi_{L-1}^*\psi_L-\mu_L p_L).\nonumber\\
\eeA
Fig. \ref{fig:chain} shows the  heat-flux balance between  the boundary currents $\Phi_jT_j$ $(j=1,L)$ and the  average
bulk flux $\overline{j^q}=1/L\sum_m j^q_m$ near a nonequilibrium stationary state with different boundary temperatures.
When the stationary state is  reached,  a relation analogous to
Eq.~(\ref{eq:udot}) holds separately at the rightmost boundary. This regime corresponds to a linear
temperature profile along the chain (see the inset) and a flat profile of $j_m^q$ (data not shown).

\section{dynamics of a dimer}%
For a better physical insight, and to appreciate the role of coupling on transport,
we now discuss the simplest realisation of the DNLS consisting
of only two coupled oscillators $L=2$ (see Fig.\ref{fig:figure0}(c) for a cartoon). 
The system is described by the non-Hermitian Hamiltonian
\beA\label{eq:hamilton}
\mathcal{H}&=&(1+i\alpha)[\omega_1(p_1)p_1+\omega_2(p_2)p_2+A_{12}\psi_1\psi^*_2\nonumber\\
           &&+A_{21}\psi_1^*\psi_2]+i\alpha\mu_1p_1+i\alpha\mu_2 p_2    
\eeA
The quantities $\omega_m(p_m)=\omega^0_m+Qp_m$ and $\alpha\omega_m(p_m)$, $m=1,2$ are respectively the non-linear frequency and damping with $Q$ the nonlinearity coefficient, while $\mu_m$ is the chemical potential. 
For simplicity we do not write the explicit dependence of the frequencies on the powers.
The coupled equations of motion, given by $\dot{\psi_m}=i\der_m^*\mathcal{H}+\xi_m$, $m=1,2$ read 

\beA
\dot{\psi}_1 &=& (i-\alpha)(\omega_1\psi_1+A_{12}\psi_2)+\alpha\mu_1\psi_1+\xi_1\label{eq:osc1}\\
\dot{\psi}_2 &=& (i-\alpha)(\omega_2\psi_2+A_{21}\psi_1)+\alpha\mu_2\psi_2+\xi_2\label{eq:osc2}
\eeA

From the previous section, one has the following expressions for particle and energy currents:
\beA
j^p_{12}=2{\Im}\average{A_{12}\psi_1^*\psi_2},\label{eq:jp}\\
j^E_{12}=2{\Re}\average{A_{12}\psi_1^*\dot{\psi}_2}\label{eq:jE}.
\eeA

When the two reservoirs have different temperatures and/or chemical potentials or an asymmetric coupling, the system reaches a non-equilibrium steady state where the currents
are constant. Thermal equilibrium, which corresponds to the case where the currents are zero is obtained where both baths have the same temperature and chemical potentials and the coupling is 
symmetric, $A_{12}=A_{21}\equiv A$. Note that if the coupling is symmetric, one has $j_{12}^{p/E}=-j_{21}^{p/E}$. However, for an asymmetric coupling those currents are different and
transport is described by the net currents $j_{net}^{p/E}=j_{12}^{p/E}-j_{21}^{p/E}$.

As discussed previously, the $I$ and $R$ components of the thermodynamical forces $F_m^{I/R}=i\der_m^*\mathcal{H}^{I/R}$, are the ones that change (resp. do not change) sign upon the time reversal operation
$\mathcal{H}^{I/R}(t)\rightarrow\mathcal{H}^{I/R*}(-t)$. To separate the Hamiltonian in $I/R$ parts, it is convenient to split the coupling between the oscillators as $A = B + C$, respectively
into Hermitian and anti-Hermitian parts. A straightforward calculation gives
\beA
\mathcal{H}^I &=&-i\alpha(\omega_1p_1+\omega_2p_2+\mu_1p_1+\mu_2p_2)\nonumber\\
		      &&-i\alpha(B_{12}\psi_1^*\psi_2+B_{21}\psi_1\psi_2^*)\nonumber\\
		      &&+C_{12}\psi_1^*\psi_2+C_{21}\psi_1\psi_2^*,\label{eq:hirr}\\
\mathcal{H}^R &=&\omega_1p_1+\omega_2p_2+B_{12}\psi_1^*\psi_2+B_{21}\psi_1\psi_2^*\nonumber\\
                       &&-i\alpha(C_{12}\psi_1^*\psi_2+C_{21}\psi_1\psi_2^*),\label{eq:hrev}
\eeA
and the thermodynamical forces read

\beA
F_1^I &=& -\alpha(\mu_1\psi_1+\omega_1\psi_1+B_{12}\psi_2)-iC_{12}\psi_2,\label{eq:firr}\\
F_1^R &=&-i(\omega_1\psi_1+B_{12}\psi_2)-\alpha C_{12}\psi_2.\label{eq:frev}
\eeA
Note that one has the same decomposition if the coupling matrix $A$ is real, but in this case $B$ and $C$ are respectively its symmetric and anti-symmetric components.
One can see here that the presence of anti-Hermitian (or anti symmetric) components adds extra terms in both the irreversible and reversible forces.

Following Eq.(\ref{eq:entropy5}), the entropy production for the dimer finally reads:
\beA
\Phi &=& 2\frac{\average{|F_1^I|^2}}{\alpha T_1}+2\frac{\average{|F_2^I|^2}}{\alpha T_2}\nonumber\\
     &&+2{\Re}\average{\der_1F_1^I}+2{\Re}\average{\der_2F_2^I},
\eeA
with $\der_mF_m=-\alpha(\mu_m+\omega_m)$.

\begin{figure}
\begin{center}
\includegraphics[width=8.0cm]{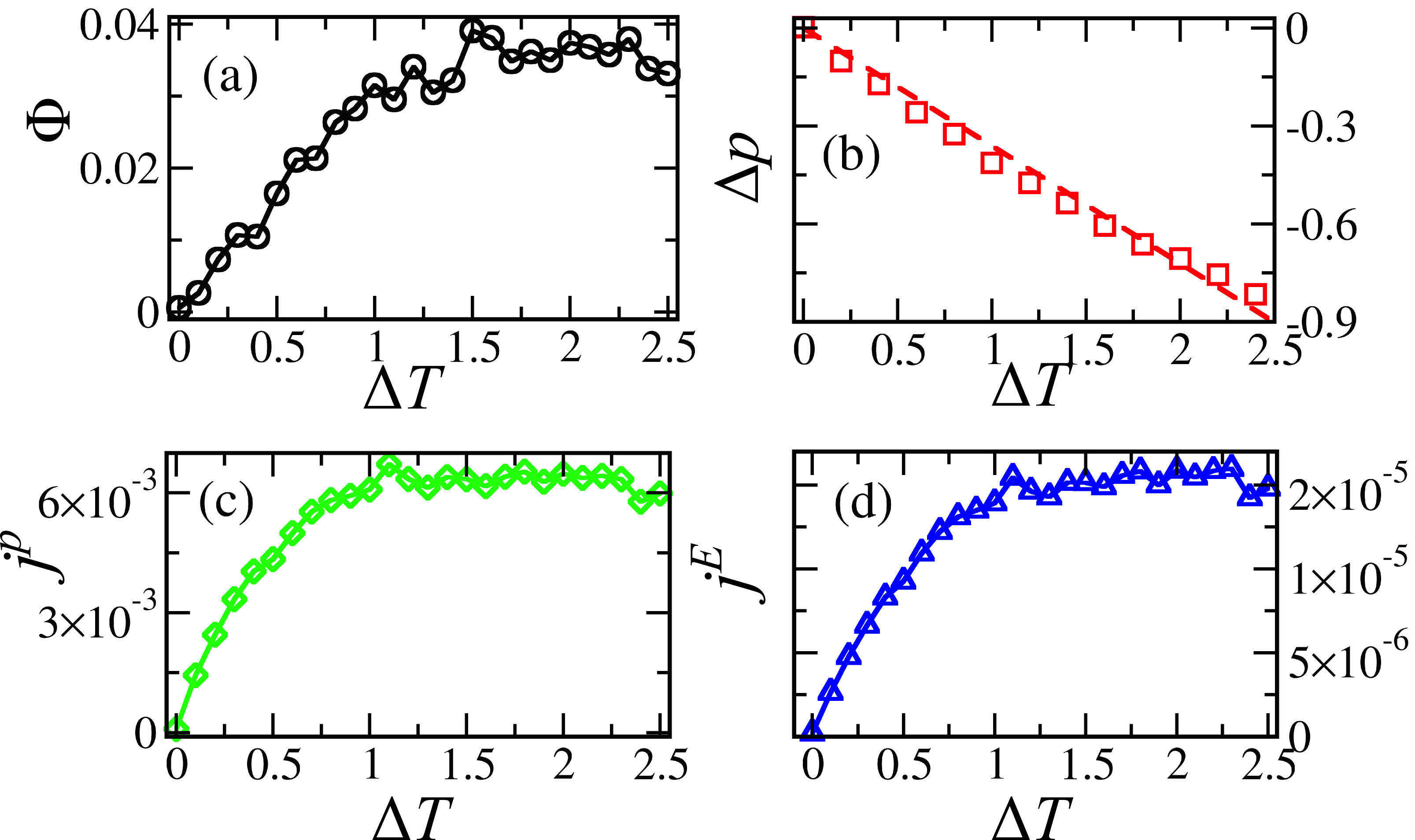}
\caption{(Color online) DNLS dimer: time-averaged observables 
as a function of $\Delta T=T_1-T_2$. Panel a) Shows the entropy flow, panel b) the power difference $\Delta p=p_1-p_2$ while panels c) 
and d) display respectively particle and energy currents. The solid lines are guides to the eye, while the dashed line in panel b) is a linear fit.
Eqs.(\ref{eq:osc1}) and (\ref{eq:osc2}) have been integrated numerically using a fourth order Runge-Kutta algorithm.
The integration has been performed for $4\times 10^6$ steps, with a time step of $10^{-3}$ model units. 
The observables where time averaged and then ensemble averaged on 64 different realisations of the thermal field.
}
\label{fig:figure1}
\end{center}
\end{figure}

We turn now to numerical simulations of Eqs.(\ref{eq:osc1}) and (\ref{eq:osc2}). In the 
following, the parameters  $\alpha=0.02$, $\omega_1^0=\omega_2^0=1$ where used.
At first, we have calculated the observables for a system with \emph{symmetric} coupling $A_{12}=A_{21}\equiv A=0.1$, keeping $T_1=0.2$ and varying $T_2$ between $0.2$ and $2.7$ model units.
Fig.\ref{fig:figure1} shows the observables as a function of $\Delta T=T_1-T_2$. One can see that both the entropy production and the currents increase linearly at low temperature
and then saturate. This behavior is similar to what has been observed in several systems previously studied, such as the spin-caloritronics diode and artificial spin chains 
\cite{borlenghi14a,borlenghi14b,borlenghi15a,borlenghi15b}. It is due to the fact that at increasing temperature, thermal fluctuation hinder synchronisation between the oscillators thus 
reducing the currents. 

The power difference $\Delta p=p_1-p_2$ decrease linearly as a function of $\Delta T$, since $p_1$ remains constant and $p_2$ is proportional to the temperature $T_2$.

\begin{figure}
\begin{center}
\includegraphics[width=8.0cm]{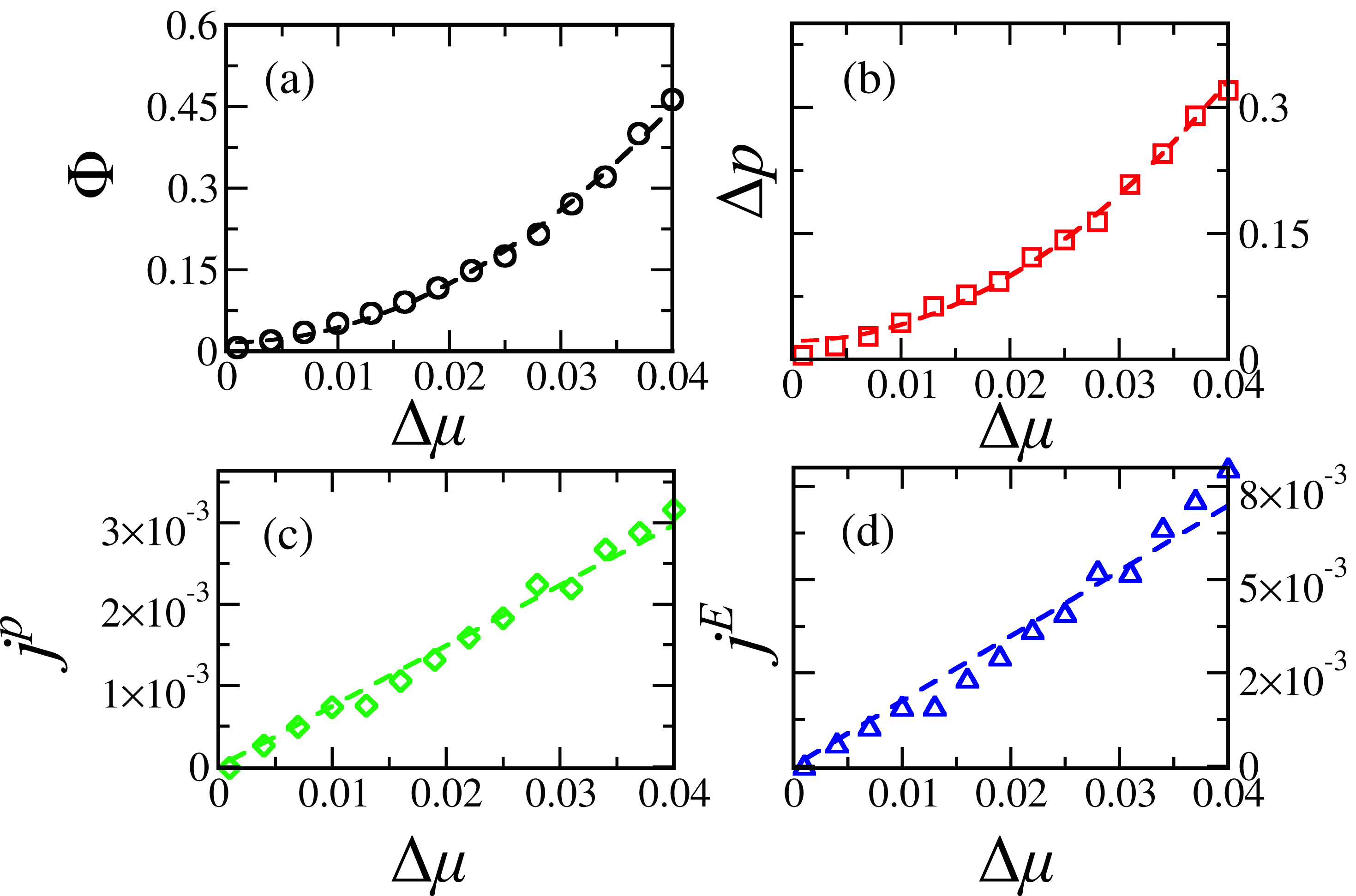}
\caption{(Color online) Time averaged observables computed as a function of the chemical potential difference $\Delta\mu=\mu_1-\mu_2$
Panel a) and b) shows respectively the entropy flow and the power difference $\Delta p=p_1-p_2$, while panels c) and d) display 
respectively the particle and energy currents.
Dashed lines in panels  a) and b) are quadratic fits, dashed lines in panels c)  and d) are linear fits.
}
\label{fig:figure3}
\end{center}
\end{figure}
Next, we focus on the effect of chemical potential difference on transport. 
In Fig.(\ref{fig:figure3}) the observables as a function of  $\Delta\mu=\mu_1-\mu_2$ are reported.
The simulations where performed keeping $T_1=T_2=0.1$ and $\mu_2=0.01$ fixed 
and varying $\mu_1$ between $0.01$ and $0.05$. One can observe that both $\Phi$ and $\Delta p$ grows quadratically, while the currents increase linearly as a function of $\Delta\mu$. Note
in particular that no saturation is observed in this case.

\begin{figure}
\begin{center}
\includegraphics[width=8.0cm]{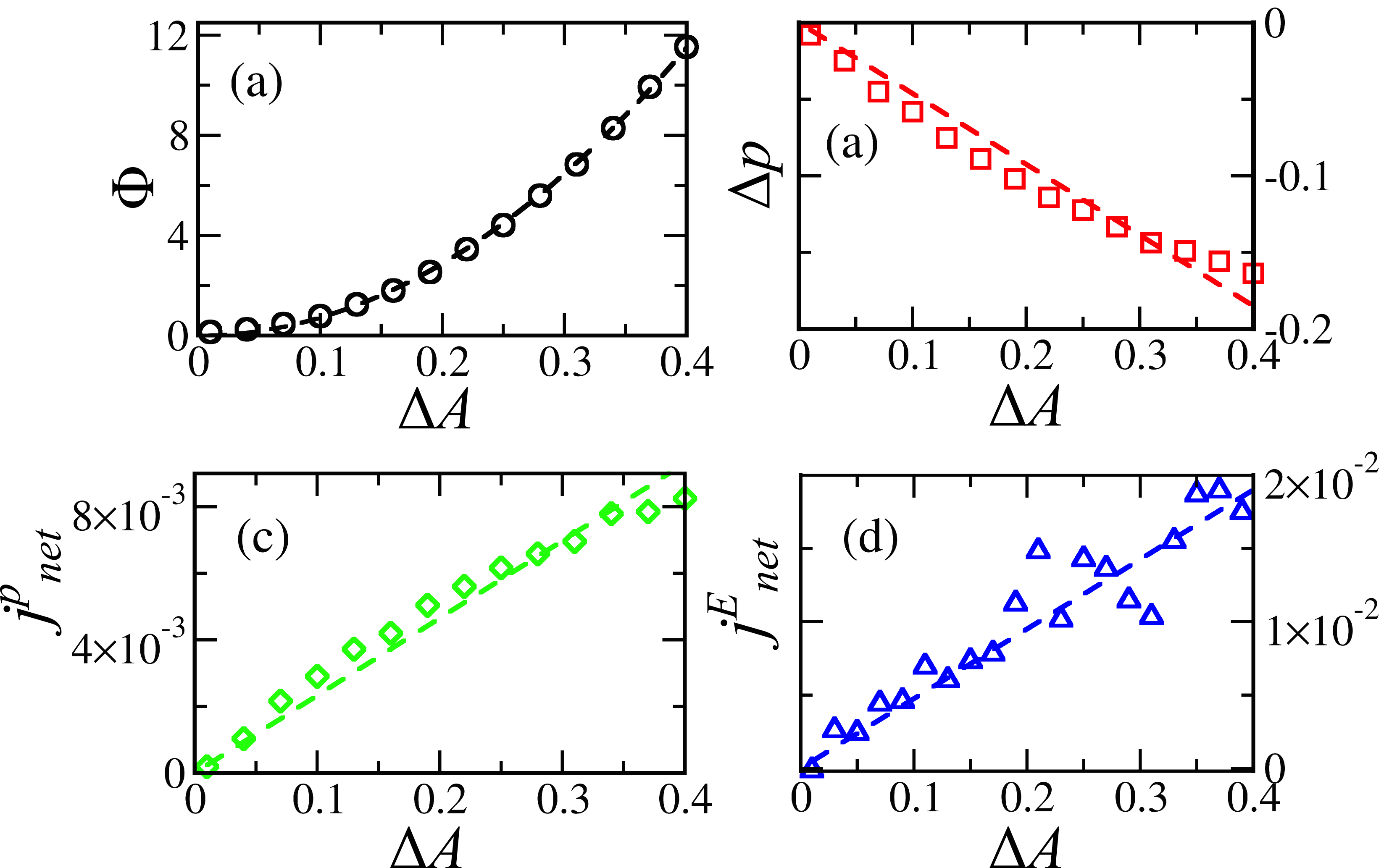}
\caption{(Color online) Time averaged observable at constant temperature, computed as a function of the coupling difference $\Delta A=A_{12}-A_{21}$, keeping $A_{21}=0.1$ fixed and increasing $A_{12}$. Panel a) and b) shows 
respectively the entropy flow and the power difference $\Delta p=p_1-p_2$ while panels c) and d) display respectively the particle and energy currents.
The Dashed line in panel  a)  is a quadratic fit, dashed lines in panels b), c)  and d) are linear fits.
}
\label{fig:figure2}
\end{center}
\end{figure}
Finally, let us discuss the case in which the model is brought outside equilibrium by
an asymmetric coupling. 
Fig. \ref{fig:figure2} displays the observables at constant temperature $T=0.2$ as a function of the asymmetry $\Delta A=A_{12}-A_{21}$ of the coupling. One can see in panel a) that the entropy flow increases quadratically with the coupling, while the
other observables are linear in $\Delta A$. Note also that the observables, and in particular the entropy production and the energy current, are much larger than in the case of symmetric coupling 
and temperature difference,
showing that the asymmetric coupling is a very efficient means to drive the system out of equilibrium.

\section{conclusions}%
In summary, we considered assembly of coupled nonlinear oscillators coupled to 
Langevin baths. Within the ST approach we compute explicit expressions for 
the entropy production rate and demonstrated their concrete use for specific model
cases: the DNLS chain and dimer. In the case of the chain, we showed how 
the approach to the steady state can be studied by monitoring $\Phi$.
For the dimer, we emphasized the role of asymmetry in the coupling as a means to effectively drive the system out of equilibrium. The asymmetry reflects the presence of anti-Hermitian components
in the Hamiltonian. 

The role of non-Hermitian Hamiltonians in classical and quantum oscillators has been long investigated \cite{dekker81}. Recently, the differences between the Lindblad and non-Hermitian formulation of
open quantum systems have been clarified \cite{sergi14}. The present work can serve to elucidate how anti-Hermitian components contribute to drive out of equilibrium this kind of systems. Generalising
these results to the case of multiplicative noise should allow to treat genuinely quantum systems and provide a connection with the formalism of quantum state diffusion equations \cite{gisin92}. 

The importance of the dimer is that it is simplest object that can be investigated, and yet it exhibits a rich dynamics due to the fact that it has two conserved quantities with associated currents.
In magnetic system and in particular in spin valve structures, the dipolar interaction between layers introduces naturally an asymmetric coupling \cite{naletov11}, and further investigation is needed to understand coupled transport 
in those systems. Most of the times these setup can be described by simple dimer models as the one treated in this paper \cite{slavin09}.

Generally speaking, the off-equilibrium observables are of importance to quantify irreversibility in a multitude of physical systems. 
Possible applications include the description of transport in mechanical oscillators \cite{salerno14}, synthetic gauge fields \cite{celi14} and topological insulators \cite{rivas16}. 
Similar expression for entropy productions have also been obtained in the context of granular media \cite{gradenigo12}.

We remark that the role of asymmetric coupling in the dynamics of oscillator network has attracted a certain attention in recent years, especially in connection with synchronisation phenomena and the dynamics of neural network
\cite{yamada03,blasius05,belych06,zeitler09,cantos16}.  Our work moves a step forward by addressing the off-equilibrium thermodynamics of those type of systems using a very general approach.

We mention also that a possible mechanism to create an asymmetric or complex coupling consists in forcing parametrically the coupled oscillators in such a 
way that the forcing has a fixed phase. 

\acknowledgements
We thank V. Tosatti and M. Polettini for illuminating discussions. This research was supported by the Stiftelsen Olle Engkvist Byggm\"astare.


\end{document}